# Why exomoons must be rare?

Rosaev A.E.

Research and Educational Center "Nonlinear Dynamics",
Yaroslavl State University, Yaroslavl, Russia E-mail:Hegem@mail.ru

The problem of the search for the satellites of the exoplanets (exomoons) is discussed recently. There are very many satellites in our Solar System. But in contrary of our Solar system, exoplanets have significant eccentricity. In process of planetary migration, exoplanets can cross some resonances with following growth of their orbital eccentricity. The stability of exomoons decreases, and many of satellites were lost. Here we give a simple example of loss satellite when eccentricity increased.

Finally, we can conclude that exomoons must be rare due to observed large eccentricities of exoplanets.

## Introduction

Despite the discovery of thousands of exoplanets and significant efforts to find, no exomoons have been detected so far [1].

A survey of 41 Kepler Objects of Interest (KOIs) for exomoons finds no compelling evidence for exomoons although thirteen KOIs yield spurious detections driven by instrumental artifacts, stellar activity and/or perturbations from unseen bodies. [3] The light curve of 1SWASP J140747.93-394542.6, can be interpreted as the transit of a giant ring system that is filling up a fraction of the Hill sphere of an unseen secondary companion, J1407b. [4]

There are few methods suggested to exomoon search for. New method to determine an exomoon's described and tested in **[1].** Two methods to determine an exomoon's suggested in **[2].** Simulations show that the required measurements will be possible with the European Extremely Large Telescope (E-ELT). Both new methods can be used to probe the origin of exomoons, that is, whether they are regular or irregular in nature. **[2]**

But how many exomoons we may expect to discover? There are large numbers of satellites are known in our Solar System. But dynamical conditions for moons are remarkable different in exoplanetary systems and in our solar system. There are few types of the satellites instabilities occur in exoplanetary systems.

Authors [5] show the dynamical instability that can happen to close-in satellites when planet oblateness is not accounted for in non-coplanar multiplanet systems. With a spherical host planet, moons within a critical planetocentric distance experience high inclinations and in some cases high eccentricities, while more distant moons orbit stably with low inclinations and eccentricities, as expected. Instability occurs while the nodal precession of the satellite and the central star (as seen from the host planet's frame) approaches the 1:1 secular resonance.[5]

The instability of close-to-planet exomoon due to *evection resonance* perturbation is described in [6]. According their results, orbits of exomoons close to planets (within 10 planet radius) cannot have long lifetime. Here we give a very simple example of the other type of exomoon's instability.

## The idea description

Let to consider planar three body problem (RTBP). Geometry of problem is given in fig.1. Main notations are: $m_1$ – main body or reduced mass in barycentre, $m_2$ – secondary mass $m_3$ – infinitesimal test particle, $f$ – is a gravity constant, $V_i$ – respective velocities, $\Delta_{ij}$ - distances between points.

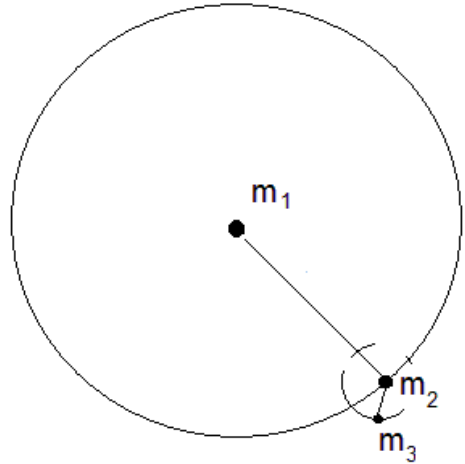

Fig.1. Problem geometry.

Let to consider a limit case $m_3=0$. Hill sphere method is well known for studying stability of asteroid small satellite. If the mass of the smaller body (e.g. planet) is $m$, and it orbits a heavier body (e.g. Sun) of mass $M$ with a semi-major axis $a$ and an eccentricity of $e$, then the radius $r$ of the Hill sphere for the smaller body (e.g. planet or asteroid primary) is, approximately[7]:

$$r \approx a(1-e) \sqrt[3]{\frac{m}{3M}} \tag{1}$$

It means, that the stability zone for the satellite is decrease with eccentricity increasing. Moreover, the solar perturbations of satellite orbit of binary eccentric are strong

If eccentricity decrease or constant, $\Delta e(t) \leq 0$, then relative energy is negative and $m_3$ orbit forever on as a satellite around $m_2$. However, if eccentricity increase with time, $\Delta e(t) > 0$, then energy can become positive, the satellite is loosed.

Let us to consider Solar System case as an example. Secular resonances play an important role in the evolution of our Solar system. The inner edge of the belt nearly coincides with the $\nu$ 6 secular resonance which is defined by g $\approx$ g6, where g is the rate of precession of the longitude of pericenter of an asteroid and g6 is the sixth eigenfrequency of the solar system planets (approximately the rate of precession of Saturn's longitude of pericenter). During the era of planet migration, the g6 resonance swept through the present position of inner asteroid belt (semimajor axis range 2.1–2.8 AU). The g6 sweeping leads to either an increase or a decrease of eccentricity depending on an asteroid's initial orbit.

The eccentricity change depends on initial value of eccentricity and Saturn migration rate. In according with theory Malhotra [8], maximal eccentricity when secular resonance crossing estimated by expression:

$$e \approx |e \pm \delta|, \qquad \delta = \left|\varepsilon\sqrt{\frac{\pi}{\lambda\sqrt{a}}}\right| \tag{2}$$

Where $\varepsilon \approx 3.5 \times 10^{-9} \exp(2a_{g6})$ in range $2 < a_{g6} < 4$ and $10^{-13} < \lambda < 10^{-11}$ depends on planet migration rate. This result is confirmed by numeric integration. [9, 10]. For example, increasing eccentricity from $e_0 = 0.1$ up to $e_0 = 0.3 - 0.4$ was observed.

Authors studying case of massive satellite show, that estimation (1) need to be corrected and binary systems (like Earth-Moon in our Solar system) are rather less stable [11].

## Numeric experiment

In details, the method of modeling is described in [12]. We have a Sun in the center of our system, planet with Jupiter mass in initially circular orbit. We add Earth and Saturn on fixed circular orbits as addition perturbers. We start with circular orbit of massless satellite at different distance from planet. After that, we numerically integrate motion of satellite during the planet eccentricity increase with different rate.

We start from initially circular orbits of satellite. Satellite orbits strongly perturbed and obtain significant eccentricity in a short time after their formation.

Extrasolar planets have orbits with significant eccentricity and close to resonances in many cases. Both circumstances make satellite lost easier. As a possible mechanism of eccentricity increasing, the crossing of secular and mean motion resonances is supposed.

First, we choose orbit close to action sphere radius. Results are given in tables 1-2. For satellite close to action sphere boundary, in all cases we observe ejection satellite when planetary eccentricity increases. As it is seems in table 1, time of satellite ejection and final eccentricity nonlinearly depends on rate of eccentricity growth. The dependence of time of satellite ejection on rate of eccentricity growth is very close to parabolic (fig.2).

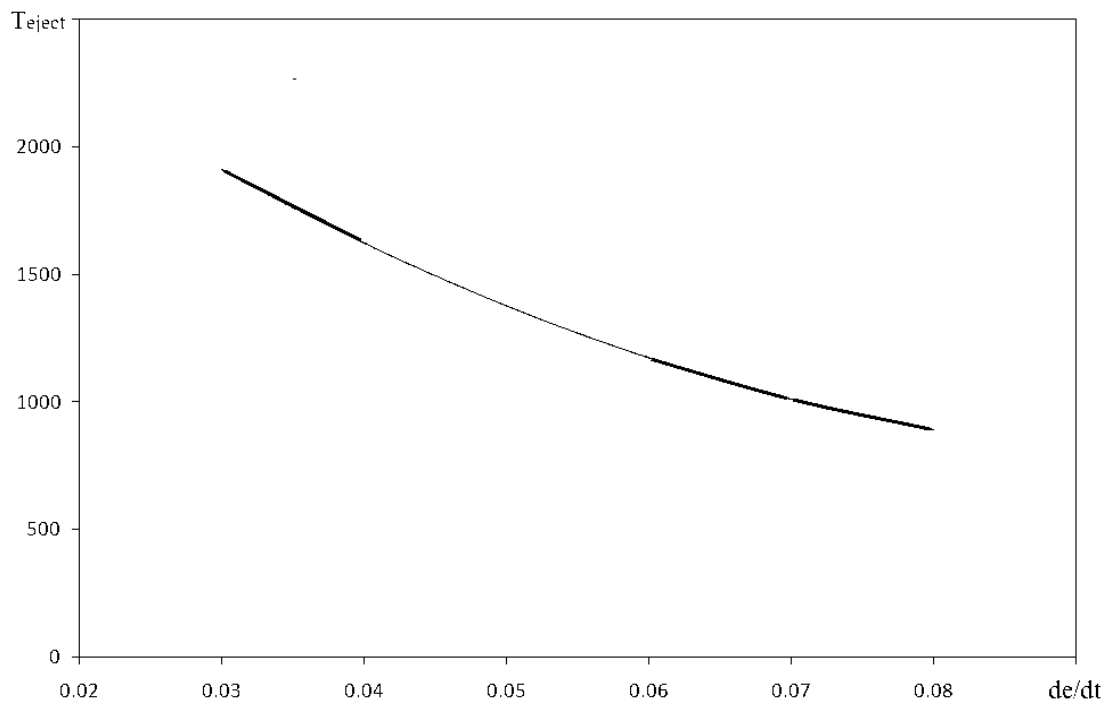

Fig.2. The dependence of satellite lifetime on eccentricity variation rate

In general, finally we have very chaotic unstable heliocentric orbit of ejected satellite. But in some cases we obtain low-eccentric heliocentric orbit of minor mass after ejection (table 2, fig 3).

Table 1. Results of modeling of satellite ejection at planetary eccentricity growth

| Orbits # | Planetocentric orbital elements | | | Planet eccentricity rate $\dot{e}$, yr$^{-1}$ | Planet eccentricity at satellite loss | Time of satellite loss |
|---|---|---|---|---|---|---|
| | $a$ R, km | e | i,° | | | |
| 2 | 20000000. | 0 | 25 | 0.03 | 0.25 | 2550 |
| 3 | 20000000. | 0 | 25 | 0.04 | 0.30 | 2320 |
| 4 | 20000000. | 0 | 25 | 0.05 | 0.23 | 1400 |
| 5 | 20000000. | 0 | 25 | 0.06 | 0.23 | 1200 |
| 6 | 20000000. | 0 | 25 | 0.07 | 0.32 | 1420 |
| 7 | 20000000. | 0 | 25 | 0.08 | 0.26 | 1000 |

Table 2. Final heliocentric orbit minor body after escape.

| Orbits # | a | e | i |
|---|---|---|---|
| 2 | 4.1076272 | 0.38365309 | 4.1381682 |
| 3 | 3.0060638 | 0.19886854 | 2.9385271 |
| 4 | 5.6227030 | 0.23397107 | 2.6866698 |
| 5 | 8.0488002 | 0.34716545 | 5.4186269 |
| 6 | 11.3134293 | 0.44063665 | 14.428551 |
| 7 | 11.0543336 | 0.58052517 | 1.9007700 |

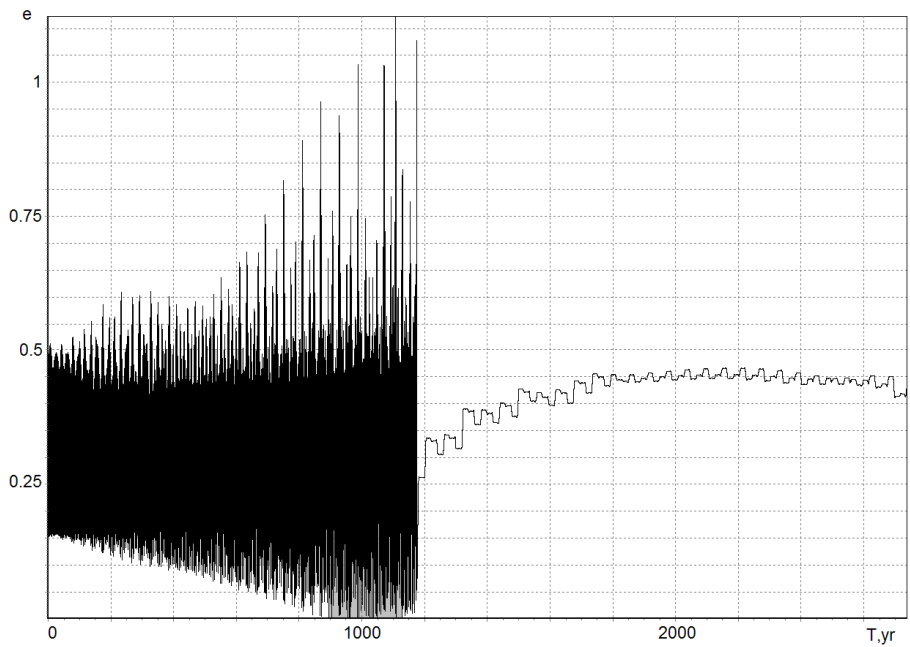

Fig. 3. The example of the satellite heliocentric eccentricity evolution at the planet eccentricity increasing

We continue our modeling with different satellite inclinations (table 3, 4). Large initial inclination satellite planetocentric orbit increases their stability, because it allows avoid close encounters with planets after ejection.

Table 3. Results of modeling of satellite ejection at planetary eccentricity growth

| Orbits # | Planetocentric orbital elements | | | Planet eccentricity rate $\dot{e}$ | Time of satellite loss | Planet eccentricity rate $\dot{e}$ yr-1 |
|---|---|---|---|---|---|---|
| | $a$ R, km | e | $i,^o$ | | | |
| 2 | 20000000. | 0 | 5 | 0.03 | 1850 | 0.00010 |
| 3 | 20000000. | 0 | 5 | 0.04 | 1750 | 0.00013 |
| 4 | 20000000. | 0 | 5 | 0.05 | 1440 | 0.00016 |
| 5 | 20000000. | 0 | 5 | 0.06 | 1175 | 0.00020 |
| 6 | 20000000. | 0 | 5 | 0.07 | 1040 | 0.00023 |

Table 4. Results of modeling of satellite ejection at planetary eccentricity growth

| Orbits # | Planetocentric orbital elements | | | Planet eccentricity rate $\dot{e}$ | Time of satellite loss | Planet eccentricity rate $\dot{e}$ yr-1 |
|---|---|---|---|---|---|---|
| | $a$ R, km | e | $i,^o$ | | | |
| 2 | 20000000. | 0 | 45 | 0.03 | 1940 | 0.00010 |
| 3 | 20000000. | 0 | 45 | 0.04 | 1620 | 0.00013 |
| 4 | 20000000. | 0 | 45 | 0.05 | 1175 | 0.00016 |
| 5 | 20000000. | 0 | 45 | 0.06 | 1170 | 0.00020 |
| 6 | 20000000. | 0 | 45 | 0.07 | 1010 | 0.00023 |
| 7 | 20000000. | 0 | 45 | 0.08 | 870 | 0.00026 |

After that, we continue calculations with satellite on initial orbit deeply inside planets Hill sphere. In this case ejection of satellite is not took place, but perturbations of initially circular satellite orbit are very significant.

During of their migrations, planets in multiplanetary systems cross a few secular resonances, with increasing their eccentricity. Observed large eccentricities of exoplanets point out on their strong perturbations in the process of migrations, when number of primordial exomoons are lost. In combination with conclusion [6] about close-to planets exomoons instability and observation evidence that number of exoplanets are very close to host stars (as following, have small Hill sphere), we cannot expect a large number of exomoons.

We can note, that exoplanets with large orbital periods in single-planetary systems are most probable candidates for exomoons search for.

## Conclusions

The principal possibility of destruction a satellite system at heliocentric orbit eccentricity growth is shown in a very simple numeric model. As a possible mechanism of eccentricity increasing, the crossing of secular and mean motion resonances is supposed. Extrasolar planets have orbits with significant eccentricity and close to resonances in many cases. As a result, we cannot expect a large number of exomoons.